\documentclass[%
 reprint,
superscriptaddress,
 amsmath,amssymb,
prb,
]{revtex4-2}

\usepackage{graphicx}% Include figure files
\usepackage{dcolumn}% Align table columns on decimal point
\usepackage{bm}% bold math
\usepackage{hyperref}
\hypersetup{hypertex=true,
	colorlinks=true,
	linkcolor=blue,
	anchorcolor=blue,
	urlcolor=blue,
	citecolor=blue}
	
\begin{document}

\preprint{APS/123-QED}

\title{Reversal of Orbital Hall Conductivity and Emergence of Tunable Topological Quantum States in Orbital Hall Insulator}% Force line breaks with \\
%\thanks{A footnote to the article title}%

\author{Shilei Ji}
\affiliation{Institute of Advanced Materials (IAM), Nanjing University of Posts and Telecommunications (NJUPT), Nanjing 210023, China.}%Lines break automatically or can be forced with \\
\author{Chuye Quan}
\affiliation{Institute of Advanced Materials (IAM), Nanjing University of Posts and Telecommunications (NJUPT), Nanjing 210023, China.}

\author{Ruijia Yao}
\affiliation{Institute of Advanced Materials (IAM), Nanjing University of Posts and Telecommunications (NJUPT), Nanjing 210023, China.}

\author{Jianping Yang}
\affiliation{School of Science, Jiangsu Provincial Engineering Research Center of Low Dimensional Physics and New Energy, Nanjing University of Posts and Telecommunications (NJUPT), Nanjing 210023, China.}

\author{Xing'ao Li}
\email{lxahbmy@126.com}
\affiliation{Institute of Advanced Materials (IAM), Nanjing University of Posts and Telecommunications (NJUPT), Nanjing 210023, China.}
\affiliation{School of Science, Jiangsu Provincial Engineering Research Center of Low Dimensional Physics and New Energy, Nanjing University of Posts and Telecommunications (NJUPT), Nanjing 210023, China.}
\affiliation{College of science, Zhejiang University of Science and Technology, Hangzhou 310023, China.}

%\collaboration{CLEO Collaboration}%\noaffiliation

\date{\today}% It is always \today, today,
             %  but any date may be explicitly specified

\begin{abstract}
Recent findings indicate that orbital angular momentum (OAM) has the capability to induce the intrinsic orbital Hall effect (OHE), which is characterized by orbital Chern number in the orbital Hall insulator. Unlike the spin-polarized channel in Quantum anomalous Hall insulator, the OAM is valley-locked, posing challenges in manipulating the corresponding edge state. Here we demonstrate the sign-reversal orbital Chern number through strain engineering by combing the $k \cdot p$ model and  first-principles calculation. Under the manipulation of strain, we observe the transfer of non-zero OAM from the valence band to the conduction band, aligning with the orbital contribution in the electronic structure.  Our investigation reveals that electrons and holes with OAM exhibit opposing trajectories, resulting in a reversal of the orbital Hall conductivity. Furthermore, we explore the topological quantum state between the sign-reversible OHE. 
\end{abstract}

%\keywords{Suggested keywords}%Use showkeys class option if keyword
                              %display desired
\maketitle

\section{Introduction} 
The orbital Hall effect (OHE) occurs in response to a transverse electric field, wherein carriers with orbital angular momentum (OAM) undergo longitudinal displacement, leading to an observable electrical response.\cite{RN776, RN741, RN744, RN746, RN752} It is regarded as the orbital analog of the spin Hall effect (SHE).\cite{RN737, RN734, RN733, RN778, RN742} However, unlike SHE, the OHE can be observed in materials with weak spin-orbit coupling (SOC).\cite{RN776, RN745, RN739} In contrast to the spin and anomalous Hall effects, directly detecting the accumulation of OAM faces inherent limitations, restricting the advancement of the OHE. Recent developments in experiments have revealed pronounced OHE in light metals, with validation achieved through the magneto-optical Kerr effect (MOKE).\cite{RN776} This discovery not only underscores the challenges associated with probing OAM but also marks a pivotal milestone in exploring the OHE and its topological properties.

OAM\cite{RN738, RN787, RN781, RN778}, referred to as orbital magnetic moment\cite{RN734, RN733, RN736, RN750, RN785, RN735} or orbital texture\cite{RN749, RN746, RN784}, plays a pivotal role as the source of the OHE—an observation substantiated through both experimental and theoretical avenues. In two-dimensional (2D) transition metal dichalcogenides (TMDs), the presence of non-zero OAM with opposite signs in two valleys, under the protection of $C_3$ symmetry, gives rise to a topologically protected OHE.\cite{RN734, RN749, RN738, RN751, RN730} This intriguing behavior is expressed through the topological invariant known as the orbital Chern number ($C_L$). Remarkably, carriers excited in the K valley exhibit positive OAM, moving in the opposite direction to those in the K$^{\prime}$ valley carrying negative OAM.\cite{RN744, RN738, RN575, RN635} The unique electronic motion gives rise to topologically protected quantum states, reminiscent of the purely spin-polarized currents observed in the quantum spin Hall effect.\cite{RN722, RN720}

Furthermore, within  TMDs, there exist other topological quantum states, such as the quantum anomalous Hall Effect (QAHE), which manifests specifically in ferromagnetic systems.\cite{RN592, RN520, RN557, RN593, RN642, RN686, RN694} A single spin-polarized electronic structure near the Fermi level leads to spin-polarized edge states in QAHE. Changing the magnetic moment direction in TMDs can induce a reversal of the conductive channel at the edge.\cite{RN593} However, in the context of OHE, the OAM is valley-locked, making it impossible for carriers excited in the K valley to carry negative OAM.\cite{RN734, RN736, RN749, RN635, RN597} Modifying the $C_L$ by manipulating  OAM in the valley is thus deemed unfeasible. The following issue arises: can the edge states of OHE be manipulated, and is it possible to alter the direction of motion for carriers with OAM?

Here we close this gap by using a strain engineering to investigate the interplay between the OAM and $C_L$. Combining the \(k\cdot p\) model with first-principles calculations, we show that non-zero OAM patterns can switch between the valence and conduction bands under strain. With the application of strain, the band gap within the valley undergoes a gradual closure and subsequent reopening. Concurrently, the orbital contributions on the valence band shift from magnetic quantum numbers \(\pm 2\) to 0. We prove that there are three different topological quantum states in this process, corresponding to OHE, QAHE, and OHE respectively, and their $C_L$ of $+1$, 0, and $-1$. Subsequently, we explore the transport behavior under various $C_L$. It becomes evident that the OAM patterns in both the valence and conduction bands dictate the types of charge carriers with non-zero OAM, thereby leading to the different topological phase transitions.

\begin{figure}[t]
	\centering
	\includegraphics[width=\linewidth]{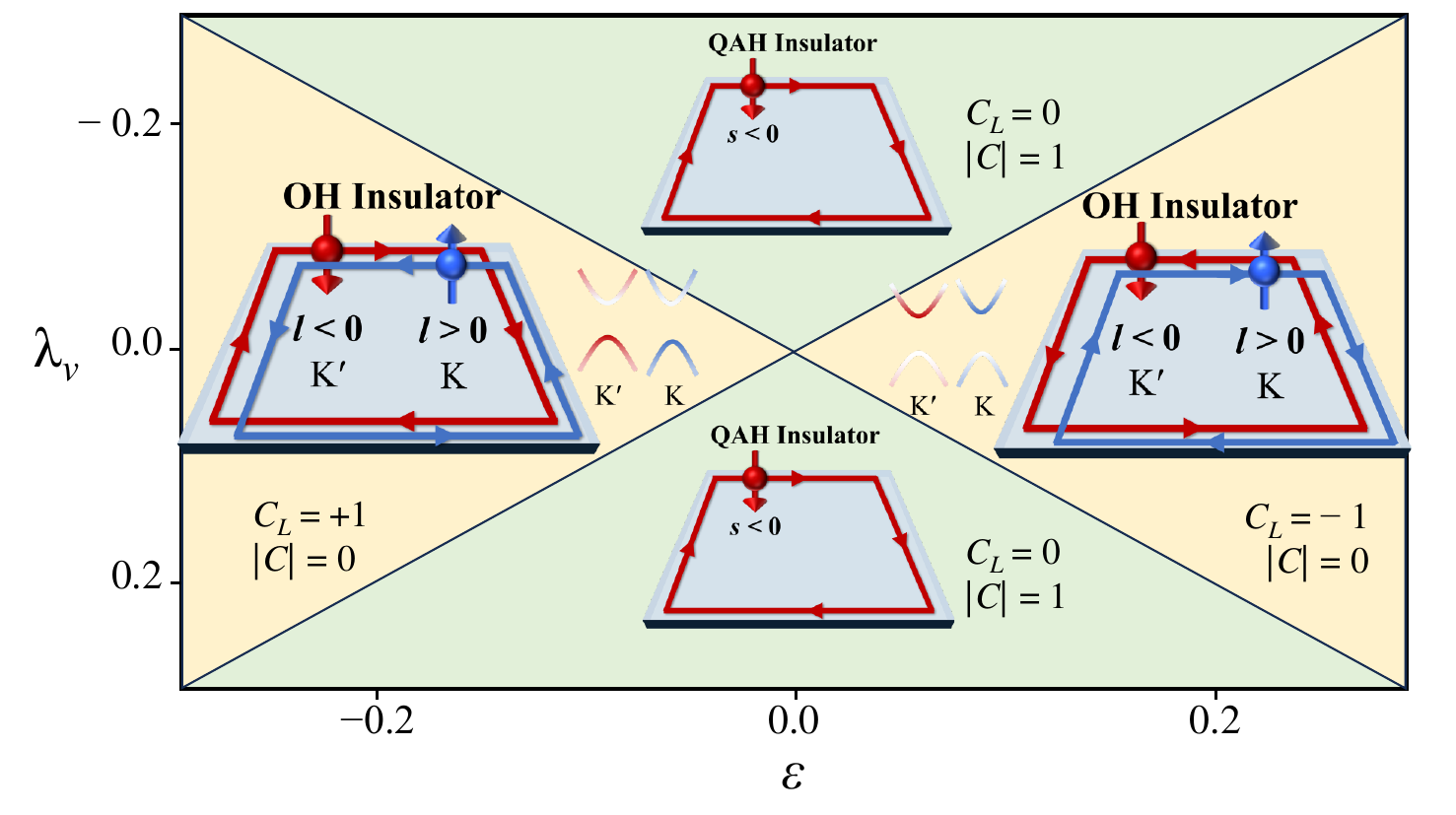}
	\caption{
	The schematic diagrams of topological quantum state calculated by two band $k \cdot p$ model.
		The other parameters are set as \(\Delta=0\) eV, \(v_F=1\) eV, and \(D=1\) eV. $s$ and $l$ are the spin and orbital angular moment, respectively. The non-zero orbital Chern number $C_L$ and Chern number $C$ represent the orbital Hall (OH) insulator and quantum anomalous Hall (QAH) insulator. With the gradual increase in the parameter $\varepsilon$, the system sequentially transitions through phases of OH insulator, QAH insulator, and back to OH insulator,  corresponding to $C_L$ of $+1$, 0, and $-1$, respectively. We depict the Orbital Angular Momentum (OAM) in the OH insulator, where the colors blue and red represent \(\pm 2\hbar\).}
	\label{fgr1}
\end{figure}

\section{Two band $k \cdot p$ analysis} 
Traditional TMD monolayers, such as H-VSe$_{2}$,\cite{RN305, RN729, RN14, RN418, RN694} breaking the spatial and time reversal symmetry, have been demonstrated to be an ideal platform to investigate OHE. The unique and clean valley model at K and K$^\prime$ points has a non-zero valley chern number at the basis of
$\psi_v^{\tau} =(|d_{x^2-y^2} \rangle + i\tau|d_{xy} \rangle) /\sqrt{2}  $ and $\psi_c^{\tau} = |d_{z^2} \rangle $, where the index $\tau= + $1 ($ - $1) represents the K (K$^\prime$) valley, and \textit{v}/\textit{c} denotes the valence/conduction state. The corresponding effective Hamiltonian $H_{\rm{eff}}$ can be written as below,\cite{RN508, RN564, RN636, RN440}
\begin{equation}\label{eqn-3}
	\begin{aligned}
		H_{\rm{eff}} = 
		v_F (\tau \hat{\sigma}_xk_x+\hat{\sigma}_yk_y) +\dfrac{\Delta}{2} \hat{\sigma}_{z} - \tau \hat{s}_{z} \lambda_v \frac{\hat{\sigma}_{z}-1}{2}.
	\end{aligned}
\end{equation}

Here, $v_F$ is the massless Fermi velocity of the Dirac electrons, $\hat{\sigma}_i$ and $\hat{s}_i$ $(i = 0, x, y, z)$ are Pauli matrices for pseudospin and spin, respectively. In addition,  the band gap of the system is represented by $\Delta$, and $\lambda_v$ is the SOC parameter for the valence band. 
The interface engineering is an effective method to manipulate the electrical properties of TMDs.\cite{RN641, RN592, RN520, RN598, RN593, RN229, RN694, RN58} In our $k \cdot p$ model, an additional term $H_{\varepsilon}$ has been added to Eq. 1,
\begin{equation}\label{eqn-3}
	\begin{aligned}
		H_{\varepsilon} = -\dfrac{D}{2}\varepsilon\hat{s}_0\hat{\sigma}_z
	\end{aligned}
\end{equation}
 which can apply a biaxial in-plane strain $\varepsilon = (a-a_0)/a_0$ to TMD. Where $ a $ and $a_0$ denote strained and equilibrium lattice constants, and $ D $ is the deformation potential.

The spin channels at the top of the valence band for two valleys are spin-up. The spin index is $ +1 $, and the Hamiltonian eigenvalues are expressed as: $ E_{\pm} =\frac{1}{2}[\tau \lambda_v \pm \sqrt{4v_F^2k^2+(\Delta_{\tau} - D\varepsilon)^2}] $, where $\Delta_{\tau}= \Delta-\tau\lambda_v$. In addition, $k^2$ is equal to $k_x^2+k_y^2$.
The Berry curvature (BC) $\Omega_{n}^{\tau,z}(\textbf{\textit{k}})$ and orbital Berry curvature (OBC) $\Omega_{n}^{\tau,\hat{L}_z}(\textbf{\textit{k}})$ at the valley can be written as
\begin{equation}\label{eqn-5}
	\begin{aligned}
		\Omega_{n}^{\tau,z}(\textbf{\textit{k}}) &= 
		-2\hbar^2 \sum_{n\neq n'}
		\frac{{\rm Im}\langle \psi_{n\textbf{\textit{k}}}|\hat{v}_x|\psi_{n'\textbf{\textit{k}}}\rangle 
			\langle \psi_{n'\textbf{\textit{k}}}|\hat{v}_y| \psi_{n\textbf{\textit{k}}}\rangle}
		{(E_{n'} - E_{n})^2} \\
		\Omega_{n}^{\tau,\hat{L}_z}(\textbf{\textit{k}}) &= 
		-2\hbar \sum_{n\neq n'}
		\frac{{\rm Im}\langle \psi_{n\textbf{\textit{k}}}|\hat{v}_x|\psi_{n'\textbf{\textit{k}}}\rangle 
			\langle \psi_{n'\textbf{\textit{k}}}|\hat{J}_y| \psi_{n\textbf{\textit{k}}}\rangle}
		{(E_{n'} - E_{n})^2} 
	\end{aligned}
\end{equation} 
Where $\hat{v}_i$ $(i = x, y)$ is the velocity operator along the $k_i$ direction and the OAM operator $\hat{J}_y$ is defined as $\hat{J}_y = \frac{1}{2}(\hat{v}_y\hat{L}_z+\hat{L}_z\hat{v}_y)$. $\hat{L}_z$ is the $z$ components of the OAM operator.
In the $k \cdot p$ model, there are two different descriptions to represent the OAM operator: the Bloch state orbital magnetic moment $\hat{L}_z^{tot}$\cite{RN736, RN738, RN750, RN635} and the intra-atomic approximation $\hat{L}_z^{intra}$\cite{RN737, RN734, RN744, RN746, RN745}.
Here, we focus on the second description: the intra-atomic approximation, where the source of the orbital magnetic moment neglected the intersite circulation current, which has been demonstrated to be effective and accurate in the study of OHE.\cite{RN737, RN734, RN744, RN746, RN745} Based on the basis function of 	$\psi_c^{\tau}$ and $\psi_v^{\tau}$ defined above, the OAM operator can be chosen as $\hat{L}_z=\rm{diag}(0,2\hbar\tau)$ for analytical solution of OHE. The BC and OBC at special valley can be simplified as 
\begin{equation}
	\begin{aligned}
		\Omega_{n}^{\tau,z}(\textbf{\textit{k}}) &= \tau \times \Omega_{n}^{\tau,\hat{L}_z}(\textbf{\textit{k}}) \\
		&= \dfrac{2\tau v_F^2 (\Delta_{\tau} - D\varepsilon)}{(4v_F^2k^2
			+(\Delta_{\tau} - D\varepsilon)^2)^{3/2}}
	\end{aligned}
\end{equation} 
The values of BC and OBC are closely related to the band gaps and have extreme values on the two valleys.
It is interesting that, the symbol of OBC on K$^{\prime}$ ($\tau = -1$) valley is the opposite of BC, while it is the same on K ($\tau = 1$) valley. This means that, when integrating BC and OBC near the K$^{\prime}$ valley, the corresponding Chern numbers are negative to each other.  The quantum anomalous Hall (QAH) insulator has a chiral, non-dissipative spin channel at the edge of the bulk state, which has an integer multiple of Chern number. In our description, the Chern number can be written as $C=\frac{1}{2} \sum_{\tau} \tau\times sgn(\Delta_{\tau} - D\varepsilon)$. In addition, the orbital Hall (OH) insulator is a topological insulator whose currents across two valleys carry opposite OAM. The orbital Chern number $C_{\hat{L}_z}=\frac{1}{2} \sum_{\tau} sgn(\Delta_{\tau} - D\varepsilon)$ is a topological invariant to characterize OH insulators. 

\begin{figure}[t]
	\centering
	\includegraphics[width=\linewidth]{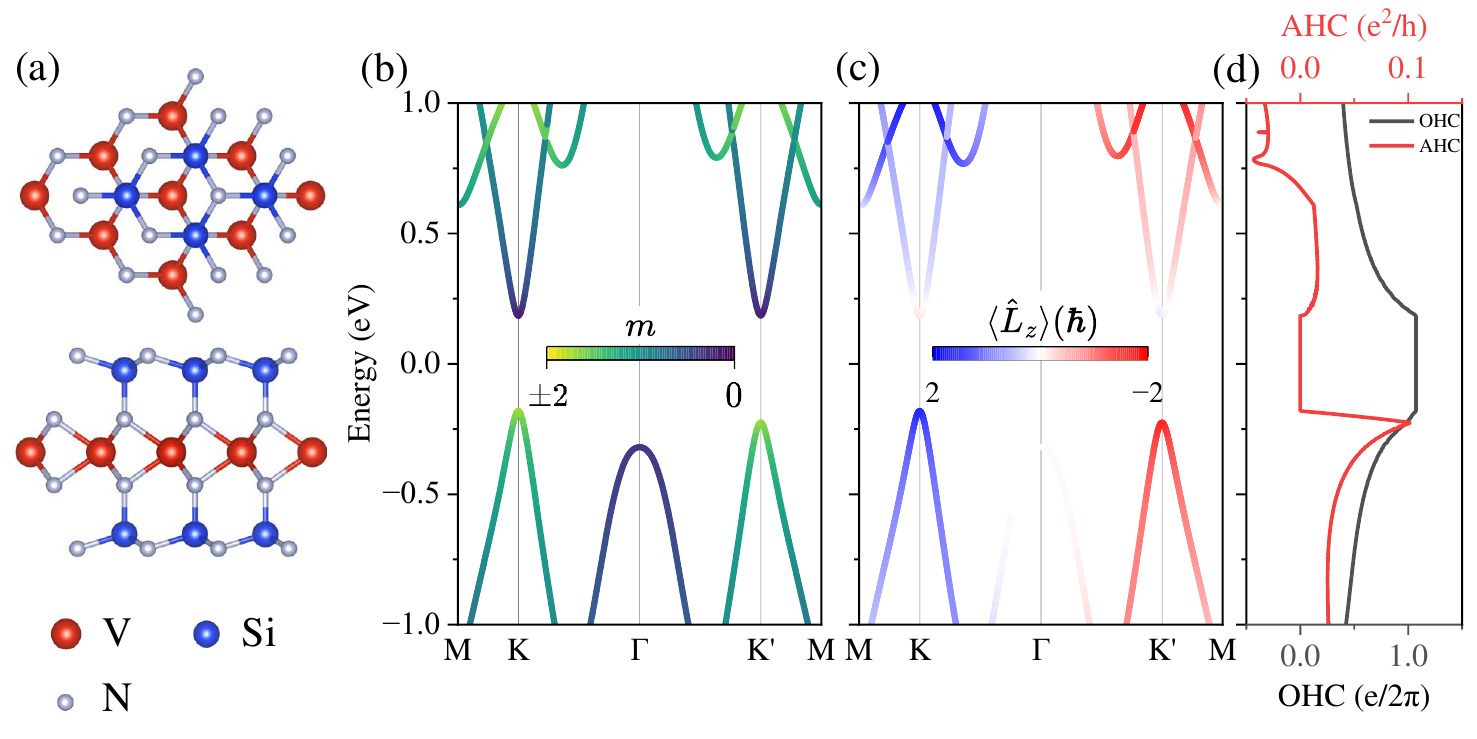}
	\caption{ 
		The first-principle calculations of 1L VSi$_{2}$N$_{4}$.
		(a) The atomic structure of 1L VSi$_{2}$N$_{4}$. Up and down panels are the top and side view, respectively. (b) The electronic and (c) orbital band structure for 1L VSi$_{2}$N$_{4}$, where $m$ and $\langle \hat{L}_{z}\rangle $ represent the magnetic quantum number and orbital angular moment. Here, $m=\pm2$ is used to represent the $d_{xy}$ and $d_{x^2-y^2}$ orbitals, while  $m=0$  is employed to  signify the $d_{z^2}$ orbital. The anomalous Hall conductivity and orbital Hall conductivity are plotted in (d), which are represented by red and black lines, respectively.
	}
	\label{fgr2}
\end{figure}

We plot a topological quantum diagram of the SOC parameter $\lambda_v$ as a function of biaxial strain $\varepsilon$ in \hyperref[fgr1]{Fig. \ref{fgr1}}. For a system without considering $\lambda_v$ and $\varepsilon$, both valleys simultaneously close, exhibiting a topologically trivial state. The application of strain can open the band gap, giving rise to the OHE with a topological invariant $C_L = \pm 1$. This indicates that the generation of OHE does not require SOC, aligning with the conclusions of previous studies.\cite{RN749, RN776, RN746} Subsequently, when $\lambda_v \neq 0$ without strain, the system is a QAH insulator with $|C| = 1$. When $|\varepsilon| > |\lambda_v|$, the system transitions from QAHE to OHE with valley polarization. This corresponds to a topological index $C_L = \pm 1$. Interestingly, for $\varepsilon < 0$, $C_L = +1$, while for $\varepsilon > 0$, \(C_L = -1\). These two types of OHE with opposite topological indices are intriguingly associated with the sign of $\varepsilon$. To investigate the change in $C_L$, we calculate the OAM in the valley,
\begin{equation}
	\begin{aligned}
		\langle \hat{L}_{z,\textbf{k}} \rangle = 
		\langle \psi_{\textbf{\textit{k}}}| \hat{L}_{z,\textbf{k}} | \psi_{\textbf{\textit{k}}}\rangle.
	\end{aligned}
\end{equation} 
The expectation values of OAM in both valleys are $\pm 2\hbar$. Notably, under the condition of $\varepsilon < 0$, $\langle \hat{L}_{z} \rangle =\pm 2\hbar$ occurs at the top of the valence band, while for $\varepsilon > 0$, it is observed at the bottom of the conduction band. This observation suggests that, while maintaining a constant OAM, strain engineering has the capability to manipulate the edge states of electrons and holes in OH insulators. The $C_L$ emerges as an effective descriptor delineating these distinct edge states.

\begin{figure}[t]
	\centering
	\includegraphics[width=\linewidth]{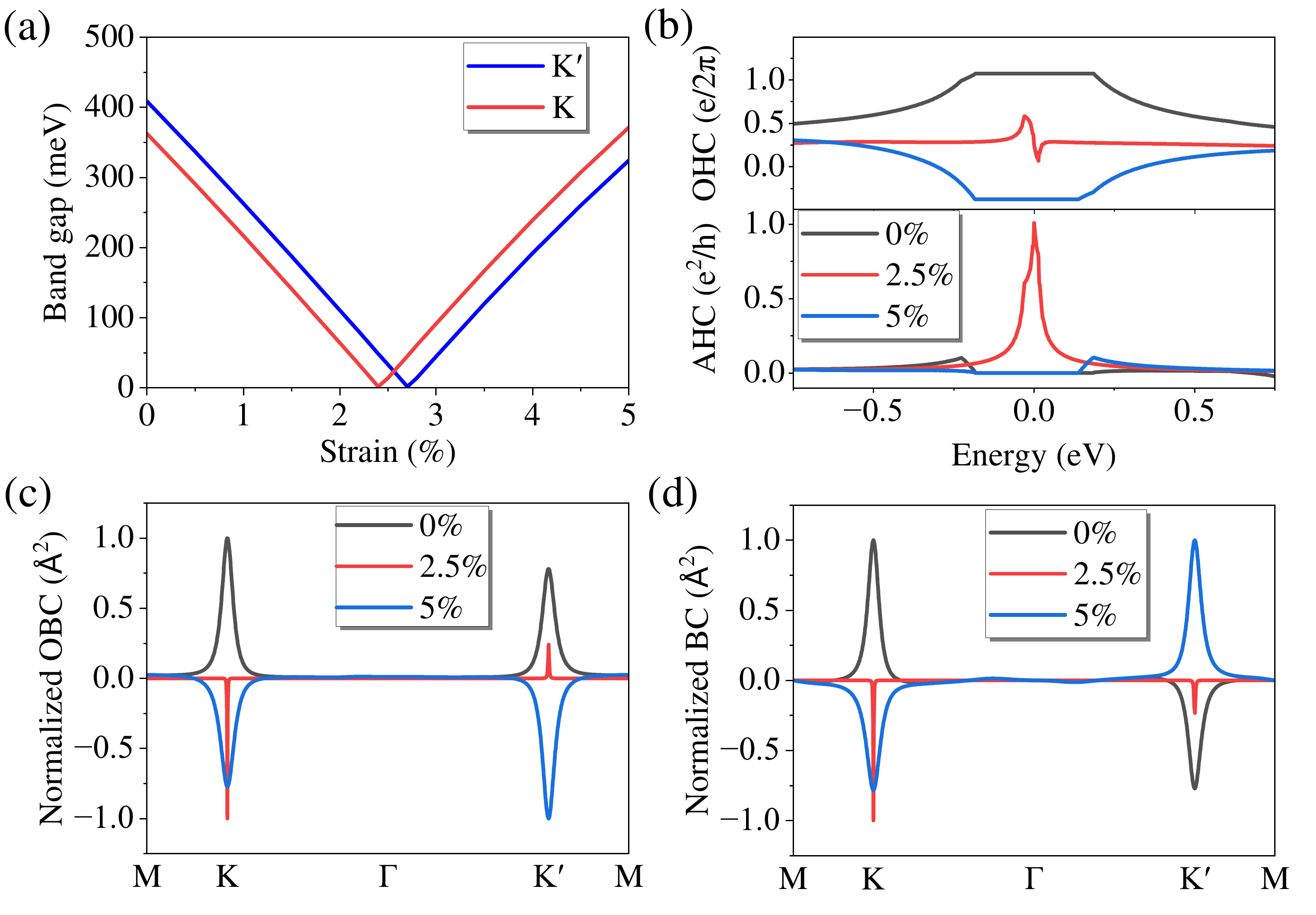}
	\caption{ 
		The transport property of 1L VSi$_{2}$N$_{4}$   with the first-principle calculations combining Wannier function.
		(a) The band gaps at two valleys as a function of strain. The Blue and red lines are the K and K$^{\prime}$ valleys. There is a band reversal near the strain of 2.5\%. We choose 0\% (black line), 2.5\% (red line), and 5\% strain (blue line) as three different cases to investigate the topological quantum state. The corresponding orbital Hall conductivity (OHC) and anomalous Hall conductivity (AHC) are plotted in (b). In addition, we analysis (c) the Normalized orbital Berry curvature and (d) Berry curvature through the first Brillouin  Zone.
	}
	\label{fgr3}
\end{figure}

\section{DFT-based calculations} 
\subsection{OHE and OAM for VSi$_{2}$N$_4$}
To realize the tunable topological quantum states, we propose a family of OH insulator, VA$_{2}$Z$_{4}$ (A = Si, Ge; Z = N, P, As) in \hyperref[fgr2]{Fig. \ref{fgr2}(a)}.\cite{RN641, RN592, RN642, RN694} Which has been extensively employed in the investigation of both the QAHE and the Valley Hall Effect.  
Note that, despite VSi$_2$N$_4$ is not a TMD (MX$_2$), in a stoichiometric sense, it is ruled by a similar low-energy theory.
In this work, taking VSi$_{2}$N$_{4}$ as an example, we investigate the strain-dependent topological phase transition. The electronic structure plotted in \hyperref[fgr2]{Fig. \ref{fgr2}(b)} is a direct gap semiconductor with the shape of valley. Due to the non-negligible SOC in transition metals, the valley splitting of VSi$_{2}$N$_{4}$ at the two valleys occurs in the valence band occupied by $d_{xy}$ and $d_{x^2-y^2}$. In addition, in \hyperref[fgr2]{Fig. \ref{fgr2}(c)}, we display the OAM-resolved band structure by calculating the expectation value of  $\hat{L}_z$. By comparing the results of $k \cdot p$ model and DFT, we find that non-zero magnetic quantum numbers ($d_{xy}$ and $d_{x^2-y^2}$) can produce large OAMs at two valleys. At two valleys, the expectation value $\langle \hat{L}_z \rangle$ can reaches $\sim\pm 2 \hbar$. When an in-plane electric field is applied to VSi$_{2}$N$_{4}$, the carriers generated in the valley carry not only the spin angular momentum but also the OAM, which is AHE and OHE, respectively. 
	For a 2D system,  the anomalous  Hall conductivity $\sigma_{xy}^{AH}$ and orbital  Hall conductivity $\sigma_{xy}^{OH}$ can be calculated by
		\begin{equation}
			\begin{aligned}
				\sigma_{xy}^{AH} &= 
				-\frac{e^2}{\hbar} \sum_{n}
				\int_{BZ}
				\frac{d^2k}{(2\pi)^2} f_{n\textbf{\textit{k}}}
				\Omega_{n}^{z}(\textbf{\textit{k}}) \\
				\sigma_{xy}^{OH} &= 
				\frac{e}{\hbar} \sum_{n}
				\int_{BZ}
				\frac{d^2k}{(2\pi)^2} f_{n\textbf{\textit{k}}}
				\Omega_{n}^{\hat{L}_z}(\textbf{\textit{k}}).
			\end{aligned}
		\end{equation} 
	Here, $\Omega_{n}^{z}(\textbf{\textit{k}})$ and $\Omega_{n}^{\hat{L}_z}(\textbf{\textit{k}})$ are Berry curvature and orbital Berry curvature, respectively. Moreover, BZ represents the Brillouin zone and $f_{n\textbf{\textit{k}}}$ is the Fermi–Dirac distribution.
The \hyperref[fgr2]{Fig. \ref{fgr2}(d)} shows the AHE and OHE of VSi$_{2}$N$_{4}$. Within the band gap, the anomalous Hall conductivity (AHC) vanishes while the orbital Hall conductivity (OHC) occurs with the value of 1.1 e/2$\pi$. Combining with $k \cdot p$ analysis, we observe $C=0$ while $C_{\hat{L}_z}=+1$, indicating that VSi$_{2}$N$_{4}$ is an OH insulator. The giant OHC arises from $d_{xy}+d_{x^2-y^2}$ orbitals and non-zero OAM.

\subsection{Interface engineering} 
 Interface engineering, such as strain, is an effective method for tuning the electronic and orbital structures of 2D materials. In order to manipulate the topological quantum states in VSi$_{2}$N$_{4}$, we apply biaxial strain to it without altering the spatial symmetry in \hyperref[fgr3]{Fig. \ref{fgr3}(a)}. 
 In the first-principle calculations, the biaxial strain can be applied by changing the lattice constant with the equation $\varepsilon=(a-a_0)/a_0$, where $a$ and $a_0$ are the strained and equilibrium lattice constant. 
 We observe a linear trend in the change of band gap with strain, consistent with the conclusion drawn in \hyperref[fgr1]{Fig. \ref{fgr1}}.(see details on electronic structures and fitting results in the Supplemental Material\cite{RNSM}) When the strain reaches 2.4\%, the band on K point closes, and it reopens as the strain further increases. At this point, there is a band inversion, with the conduction band contributed by $d_{xy}$ and $d_{x^2-y^2}$ orbitals, while the valence band is contributed by $d_{z^2}$ orbital. When the strain reaches 2.7\%, a similar transition occurs in the K$^{\prime}$ valley.

\begin{figure}[t]
	\centering
	\includegraphics[width=\linewidth]{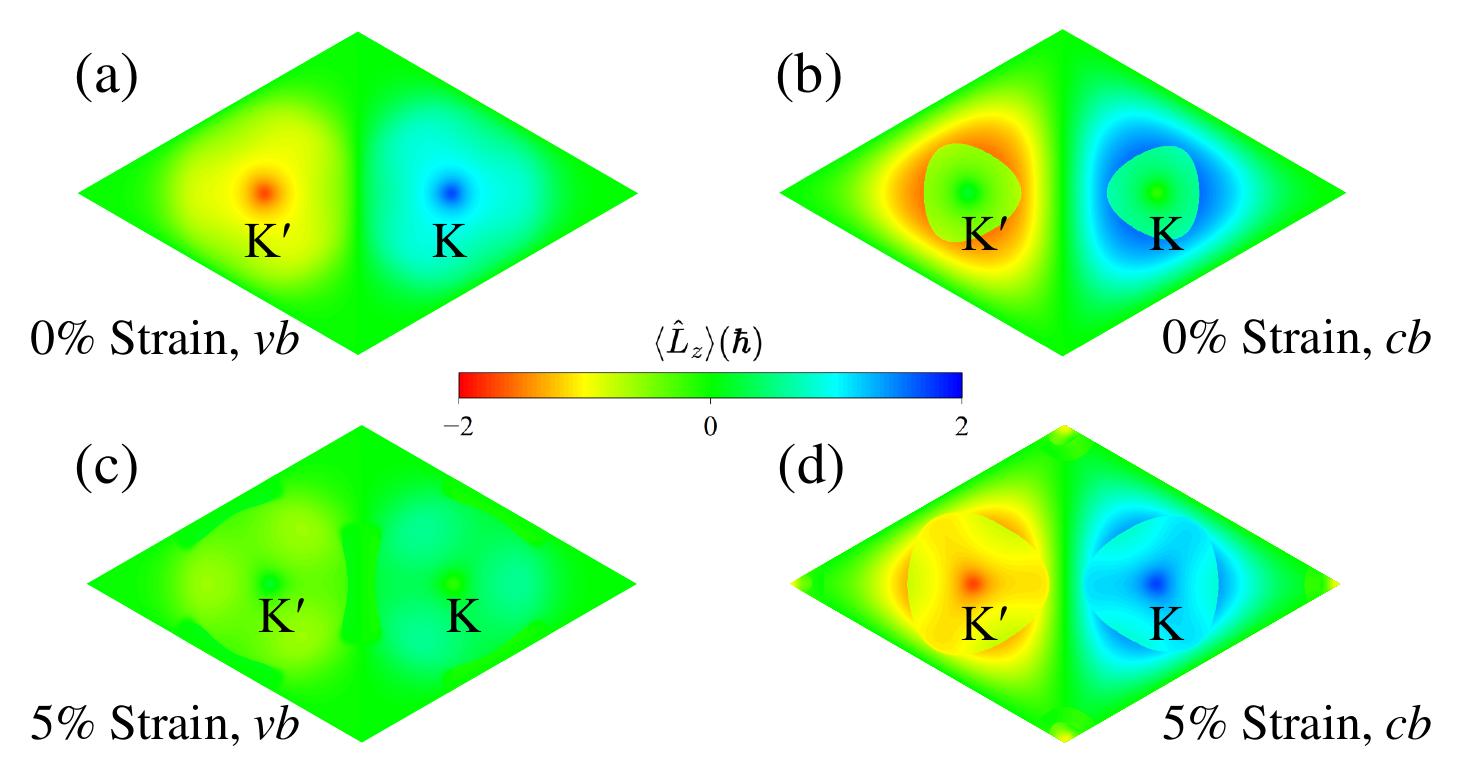}
	\caption{ 
		The orbital angular moment (OAM) distribution of (a) valance band (\textit{vb}) at 0\% strain, (b) conduction band (\textit{cb}) at 0\% strain, (c) \textit{vb} at 5\% strain, and (d) \textit{cb} at 5\% strain. The blue/red dot represents $\langle \hat{L}_z \rangle = +2/-2 \hbar$.
		The OAM distribution of 1L VSi$_{2}$N$_{4}$ is calculated by the first-principle calculations combining Wannier function.
	}
	\label{fgr4}
\end{figure}

According to the $k \cdot p$ model, we categorize VSi$_{2}$N$_{4}$ under strain into three cases, corresponding to OH, QAH, and OH insulators as outlined in \hyperref[fgr1]{Fig. \ref{fgr1}}. The detailed orbital contribution can be obtained in the Supplemental Material\cite{RNSM}. Subsequently, we calculate the AHE and OHE for the three cases in \hyperref[fgr3]{Fig. \ref{fgr3}(b)}. A significant OHC accompanying $C_L = \pm 1$ is observed when the strain is 0\% and 5\%. Since there are not only two bands contributing near the Fermi level, OHCs are not equal to zero in all three cases. 
 In particular, the OHC is about $-0.4$ $e/2\pi$  for the 5\% strain, which is mainly because the  $k \cdot p $ model can only describe the situation near the Fermi level and cannot characterize the electrical properties on all energy levels. As supported by the Fig. S4 in the Supplemental Material\cite{RNSM}, as the number of energy levels considered in OHC calculation increases, the value of OHC tends to converge. We find that the contribution of OHC in the band gap is still dominated by the two valleys, similar to the case in OBC in \hyperref[fgr3]{Fig. \ref{fgr3}(c)}. 
 By fitting the first-principles calculation with $k \cdot p $ model, we confirm that, 1L VSi$_2$N$_4$ is an OH insulator with the $C_L = -1$ at the strain of 5\%. Moreover, at 2.5\% strain, VSi$_{2}$N$_{4}$ becomes a QAH insulator with \(|C| = 1\). Tensile strain induces changes in the band structure, leading to alterations in the topological quantum states. \hyperref[fgr3]{Figs. \ref{fgr3}(b) and (c)} illustrate the BC and OBC. In the strain of 0\% and 5\%, the sign of the BC is opposite in the two valleys, leading to the disappearance of AHE. The OBC, on the other hand, exhibits a different behavior. In both cases, the OBC has the same sign, indicating that carriers with opposite OAM in the two valleys have opposite trajectories, resulting in a non-zero \(C_L\). At this point, VSi$_{2}$N$_{4}$ becomes an OH insulator. When the case switches to 2.5\% strain, the conclusions are reversed compared to the above. Therefore, under the modulation of a strain field, VSi$_{2}$N$_{4}$ exhibits rich electronic properties and topological quantum states.

\subsection{OAM for different topological quantum states} 
Although VSi$_{2}$N$_{4}$ returns to OHE under strain, it is interesting to note that the topological index is not the same as that without strain. As we can see from \hyperref[fgr3]{Fig. \ref{fgr3}(b)}, the strain changes the direction of movement of the carriers in two valleys, resulting in OHC with opposite signs. The different orbital Chern numbers, namely $-1$ and $+1$, are observed. Interestingly, the OAM in the $k \cdot p$ model is locked to the valley ($\langle \hat{L}_z^{\tau} \rangle = 2\hbar\tau$). This means that the OAM of VSi$_{2}$N$_{4}$ under strain does not change. However, the $C_L$ shifts from $+1$ to $-1$, indicating a reversal in the direction of motion for carriers carrying OAM. To understand the underlying principles, we investigate the OAM among different quantum states. Detailed OAM for the valence band and conduction band is depicted in \hyperref[fgr4]{Fig. \ref{fgr4}}. In the absence of strain, the OAM on the valence band is $\sim\pm2\hbar$, while the OAM near the conduction band valley is approximately $0\hbar$. When applying an electric field, the motion of holes on VSi$_{2}$N$_{4}$ is topologically protected due to the non-zero OAM. Subsequently, we shift our focus to a 5\% strain, where there is a reversal in the orbital contributions of the valence and conduction bands. This results in $\langle \hat{L}_z^{vb} \rangle \sim 0 \hbar$ in the valence band and $\langle \hat{L}_z^{cb}  \rangle \sim \pm 2\hbar$ in the conduction band near the valley. The electrons excited in OH insulator carry OAM of $\pm 2\hbar$ in opposite directions compared to 0\%, and this process is topologically protected. In both cases, the different transport modes of electrons and holes carrying OAM lead to a shift in the $C_L$ from $+1$ to $-1$. In addition, we summarize the OAM patterns of three cases in Fig. S3 in the Supplemental Material\cite{RNSM}. We observed distinct OAM patterns for different quantum states. Under strain engineering, the unique OAM patterns serve as a distinguishing feature for discerning the topological quantum states. Currently, directly detecting current with OAM in experiments remains challenging. Our work provides a novel perspective for exploring OHE.

%\section{Conclusion}
\section{Conclusion} 
In summary, we have established the connection between the OHE and OAM for 2D TMDs. The types of carriers with non-zero OAM can effectively influence the $C_L$ of OH insulators. Additionally, based on the band inversion under strain, we propose an effective method to manipulate the OAM and $C_L$ of the system. Taking VSi$_{2}$N$_{4}$ as an example, we demonstrate the changes in AHC and OHC under strain and observe topological phase transitions from OHE to QAHE to OHE. However, applying tensile strain does not revert VSi$_{2}$N$_{4}$ to its initial topological state, as the $C_L$ becomes opposite to that without strain. The OAM patterns in the valence and conduction bands reveal that $\langle \hat{L}_z \rangle= \pm 2\hbar$ transitions from the valence to the conduction bands. The different OAM patterns can give us a new perspective to understand and manipulate the OH insulator. In experiments, the detection and verification of intrinsic OHC and its topological invariant still face great challenges. Our work shows that distinct OAM patterns can produce different topological quantum states. Therefore, for TMDs materials, we propose such an idea to solve the above problems. By employing optical and electrical methods to detect the OAM patterns on the valleys of valence and conduction bands, we can definitively confirm the presence of the OH insulator.

\begin{acknowledgments}
	We acknowledge the fundings from National Natural Science Foundation of China (Grant No. 51872145), and Postgraduate Research \& Practice Innovation Program of Jiangsu Province (Grant No. KYCX20\_0748, KYCX19\_0935, KYCX23\_0977).
\end{acknowledgments}

\bibliography{apssamp}% Produces the bibliography via BibTeX.

\end{document}